\input{aipcheck}

\documentclass[
    ,final            
  ]
  {aipproc}

\layoutstyle{6x9}

\begin{document}

\title{Substellar companions and the formation of hot subdwarf stars}


\author{S. Geier}{
  address={Dr. Karl Remeis-Observatory \& ECAP, Astronomical Institute,
Friedrich-Alexander University Erlangen-Nuremberg, Sternwartstr. 7, D 96049 Bamberg, Germany}
}

\author{U. Heber}{
  address={Dr. Karl Remeis-Observatory \& ECAP, Astronomical Institute,
Friedrich-Alexander University Erlangen-Nuremberg, Sternwartstr. 7, D 96049 Bamberg, Germany}
}

\author{A. Tillich}{
  address={Dr. Karl Remeis-Observatory \& ECAP, Astronomical Institute,
Friedrich-Alexander University Erlangen-Nuremberg, Sternwartstr. 7, D 96049 Bamberg, Germany}
}

\author{H. Hirsch}{
  address={Dr. Karl Remeis-Observatory \& ECAP, Astronomical Institute,
Friedrich-Alexander University Erlangen-Nuremberg, Sternwartstr. 7, D 96049 Bamberg, Germany}
}

\author{T. Kupfer}{
  address={Dr. Karl Remeis-Observatory \& ECAP, Astronomical Institute,
Friedrich-Alexander University Erlangen-Nuremberg, Sternwartstr. 7, D 96049 Bamberg, Germany}
}

\author{V. Schaffenroth}{
  address={Dr. Karl Remeis-Observatory \& ECAP, Astronomical Institute,
Friedrich-Alexander University Erlangen-Nuremberg, Sternwartstr. 7, D 96049 Bamberg, Germany}
}

\author{L. Classen}{
  address={Dr. Karl Remeis-Observatory \& ECAP, Astronomical Institute,
Friedrich-Alexander University Erlangen-Nuremberg, Sternwartstr. 7, D 96049 Bamberg, Germany}
}

\author{P. F. L. Maxted}{
  address={Astrophysics Group, Keele University, Staffordshire, ST5 5BG, UK}
}

\author{R. H. \O stensen}{
  address={Institute of Astronomy, K.U.Leuven, Celestijnenlaan 200D, B-3001 Heverlee, Belgium}
}

\author{B. N. Barlow}{
  address={Department of Physics and Astronomy, University of North Carolina, Chapel Hill, NC 27599-3255, USA}
}

\author{T. R. Marsh}{
  address={Department of Physics, University of Warwick, Conventry CV4 7AL, UK}
}

\author{B. T. G\"ansicke}{
  address={Department of Physics, University of Warwick, Conventry CV4 7AL, UK}
}

\author{R. Napiwotzki}{
  address={Centre of Astrophysics Research, University of Hertfordshire, College
  Lane, Hatfield AL10 9AB, UK}
}

\author{S. J. O'Toole}{
  address={Australian Astronomical Observatory, PO Box 296, Epping, NSW, 1710, Australia}
}

\author{E. W. G\"unther}{
  address={Th\"uringer Landessternwarte Tautenburg, Sternwarte 5, D-07778 Tautenburg, Germany}
}

\begin{abstract}
We give a brief review over the observational evidence for close substellar companions to hot subdwarf stars. The formation of these core helium-burning objects requires huge mass loss of their red giant progenitors. It has been suggested that besides stellar companions substellar objects in close orbits may be able to trigger this mass loss. Such objects can be easily detected around hot subdwarf stars by medium or high resolution spectroscopy with an RV accuracy at the ${\rm km\,s^{-1}}$-level. Eclipsing systems of HW\,Vir type stick out of transit surveys because of their characteristic light curves. The best evidence that substellar objects in close orbits around sdBs exist and that they are able to trigger the required mass loss is provided by the eclipsing system SDSS\,J0820+0008, which was found in the course of the MUCHFUSS project. Furthermore, several candidate systems have been discovered. 
\end{abstract}

\maketitle


\section{Introduction}

Hot subdwarf stars (sdO/Bs) are core-helium burning stars located at the extreme blue end of the horizontal branch \citep[for a review see][]{heber09}. After leaving the main sequence the progenitors of these objects  evolve to become red giants, ignite helium and settle down on the extreme horizontal branch (EHB). Unlike normal stars, the sdB progenitors must have experienced a phase of extensive mass loss on the red giant branch to explain the high temperatures and gravities observed at the surface of hot subdwarf stars. After consumption of the helium fuel they evolve directly to white dwarfs avoiding a second red-giant phase. What causes this extreme mass loss remains an open question.  

About half of the sdB stars reside in close binaries with periods ranging from a few hours to a few days \citep{maxted01,napiwotzki04}. Because the components' separation in these systems is much less than the size of the subdwarf progenitor in its red-giant phase, these systems must have experienced a common-envelope and spiral-in phase \citep{han02,han03}. In the standard scenario, two
main-sequence stars of different masses evolve in a binary system. As soon as the more massive one reaches  the red-giant phase and fills its Roche lobe, mass is transferred to the companion star. When mass transfer is unstable, the envelope of the giant will engulf the companion star and form a common envelope. The two stellar cores lose orbital energy due to friction in the envelope and spiral towards each other until enough orbital energy has been deposited within the envelope to eject it. The end
product is a much closer system containing the core of the giant,
which then may become an sdB star, and a main-sequence companion. 

Although the common-envelope ejection channel is not yet properly understood in detail, it provides a reasonable explanation for the extra mass loss required to form sdB stars. However, for about half of all known subdwarfs there is no evidence for close stellar companions as no radial velocity variations exceeding are found. Although in some cases main sequence companions are visible in the spectra, it remains unclear whether these stars are close enough to have interacted with the sdB progenitors, because no orbital parameters have been measured yet. Among other formation scenarios, the merger of two helium white dwarfs has often been suggested to explain the origin of single sdB stars \citep{han02,han03}. 

\citet{soker98} suggested that sub-stellar objects like brown dwarfs and planets may also be swallowed by their host star and that common-envelope ejection could form hot subdwarfs. Substellar objects with masses higher than $\approx10\,M_{\rm J}$ were predicted to survive the common-envelope phase and end up in a close orbit around the stellar remnant, while planets with lower masses would entirely evaporate or merge with the stellar core (see Bear \& Soker these proceedings). The stellar remnant is predicted to lose most of its envelope and evolve towards the EHB. A similar scenario has been proposed to explain the formation of apparently single low mass white dwarfs \citep{nelemans98}. The discovery of a brown dwarf with a mass of $0.053\pm0.006\,M_{\rm \odot}$ in an $0.08\,{\rm d}$ orbit around such a white dwarf supports this scenario and shows that substellar companions can influence the outcome of stellar evolution \citep{maxted06}.

The planet discovered to orbit the sdB pulsator V\,931\,Peg with a period of $1\,170\,{\rm d}$ and a separation of $1.7\,{\rm AU}$ was the first planet found to have survived the red-giant phase of its host star \citep{silvotti07}. Serendipitous discoveries of two substellar companions around the eclipsing sdB binary HW\,Vir \citep{lee09} and one brown dwarf around the similar system HS\,0705$+$6700 \citep{qian09} followed. These substellar companions to hot subdwarfs have rather wide orbits, were not engulfed by the red giant progenitor and therefore could not have influenced the evolution of their host stars. But the fact that substellar companions in wide orbits around sdBs seem to be common  suggests that similar objects closer to their host stars might exist as well \citep[for a review see][]{schuh10}. 

\section{Methods to search for close substellar companions}

If substellar companions like planets or brown dwarfs should be able to trigger CE ejection and form sdBs, they must have been close enough to be swallowed by the red-giant progenitors. This means that the separation between star and companion in this phase must have been of the order of the red giant's radius. During the main sequence phase the companion might have migrated in from an initially wider orbit \citep{silvotti07}. Spiralling into the common envelope the orbit of the companion shrinks again. The orbital periods of the systems where the companions survived the CE ejection are predicted to range from a few hours to a few days \citep{soker98}.

Since sdB stars have masses of only about $0.5\,M_{\rm \odot}$ or less, close substellar companions cause a significant radial velocity (RV) variability of the primaries. Depending on the companion mass and the orbital period the shifts can be as high as $>50\,{\rm km\,s^{-1}}$ (see Classen et al. these proceedings). It is therefore obvious that ${\rm m\,s^{-1}}$-accuracy is not needed to search for substellar companions to sdB stars. Medium or high resolution spectrographs ($R\simeq2000-50\,000$) are perfectly suited for this purpose. No sophisticated wavelength calibration efforts (e.g. iodine cell) are necessary. Since sdBs are hot stars, features of substellar companions are usually not visible in optical spectra. Therefore only lower limits can be put on the mass of the unseen companions using the RV method. 

Another efficient way to discover close substellar companions to sdBs is searching for transits in their light curves. Hot subdwarfs are compact stars with radii ranging from $0.1$ to $0.3\,R_{\rm \odot}$. Substellar companions like hot Jupiters and brown dwarfs on the other hand are of similar size ($\simeq0.1\,R_{\rm \odot}$). If the orbital period of an sdB binary is only a few hours, the probability to find at least partially eclipsing systems is quite high and the eclipses themselves can be very deep. A combined spectroscopic and photometric analysis allows to put tight constraints on the companion mass in this case.


\section{Candidate sdB systems with substellar companions}

\subsection{Binaries with small RV variability}

Since most sdB binaries are single-lined systems, only lower limits can be put on the masses of their companions usually assuming a canonical EHB mass of $\simeq0.47\,M_{\rm \odot}$. \citet{edelmann05} discovered two binaries with minimum companion masses below the hydrogen-burning limit ($\simeq0.08\,M_{\rm \odot}$), but \citet{geier10a} showed that these binaries are most likely seen at low inclination angles and that the companions are therefore much more massive. The very short period system PG\,1017$-$086 \citep{maxted02,geier10a} on the other hand may have a brown dwarf companion with a mass exceeding $0.06\,M_{\rm \odot}$.

The smallest RV variability ever measured for an sdB binary was reported by \citet{geier09} in the case of  HD\,149382. With a period of $2.39\,{\rm d}$ and an RV-semiamplitude of only $2.3\,{\rm km\,s^{-1}}$ the most likely companion mass ranges from $8$ to $23\,M_{\rm J}$. Although these parameters are in perfect agreement with the formation scenario proposed by \citet{soker98} it is not yet clear whether this sdB is really orbited by a substellar companion. Jacobs et al. (these proceedings) took high resolution spectra and did not detect significant RV variations within tenths of days. In order to resolve this issue we obtained high resolution follow-up spectra with AAT/CYCLOPS (O'Toole et al. these proceedings). \citet{geier10b} discussed possible sources for the IR-excess in the flux distribution of HD\,149382 discovered by \citet{ulla98} including a contribution from an irradiated substellar companion. However, \citet{oestensen05} discovered a companion at a separation of about $75\,{\rm AU}$, which causes the IR-excess (see Jacobs et al. these proceedings). 

The most important setback of the RV method is the unknown inclination of the binaries. From an RV curve of a single-lined binary alone only a lower limit for the companion mass can be given. For a single object it is therefore impossible to prove the existence of a substellar companion, because a more massive stellar companion seen at low inclination looks exactly the same. Since about $50\%$ of all known sdBs are in close binary systems with stellar companions, there must be a certain number of such systems seen at low inclinations. A large sample of sdBs has to be studied to decide whether the fraction of systems with small RV variations is consistent with the low-inclination extension of the known sdB binary population or not. A higher fraction than expected would be an indication for a population of substellar companions. We have started a project to look for small RV shifts and derive orbital solutions of bright sdBs with high resolution spectroscopy (see Classen et al. these proceedings; O'Toole et al. these proceedings).

\subsection{Eclipsing systems}

Eclipsing binaries are key objects for our understanding of stellar evolution, because most parameters of the components can be constrained by a combined photometric and spectroscopic analysis. Eclipsing systems with hot subdwarf primaries and low mass stellar companions are known since more than four decades. In recent years more than ten of these systems have been discovered and about half of them have been studied in detail. The two most striking similarities between all these objects are their short orbital periods ($0.09-0.26\,{\rm d}$) and the low masses of their companions ($\simeq0.1\,M_{\rm \odot}$). 

The formation of the sdBs in these systems, named after the prototype HW\,Vir \citep{menzies86}, requires a CE phase, because the orbital periods are too short to be consistent with any other known formation scenario. Since the progenitor of the sdB star must have evolved to the tip of the RGB to ignite core helium-burning, a lower limit of about $0.8\,M_{\rm \odot}$ can be derived for the companion mass on the main sequence. Progenitors with masses as high as $2.0$ or $3.0\,M_{\rm \odot}$ may be possible as well. The initial binary must have been close enough to start unstable mass transfer in the RGB phase ($\approx50-100\,R_{\rm \odot}$). The low mass companions eventually enter the CE and trigger the ejection of the envelope. Close binary systems with such extreme mass ratios of ten or more on the main sequence are predicted to be rare. 

The fact that about $10\%$ of all known sdB binaries are in HW\,Vir systems proves that low mass stellar companions play a crucial role in sdB formation. Since the masses of these companions are close to the limit for core hydrogen-burning the question was asked whether substellar companions in HW\,Vir systems exist as well. In the cases of AA\,Dor \citep{rauch00} and HS\,2231$+$2441 \citep{oestensen08} the discovery of brown dwarf companions has been reported. However, in both cases the conclusion was based on the assumption of a particularly low sdB mass and is still under debate \citep[][M\"uller et al. these proceedings]{for10}. 

Two new HW\,Vir systems have been discovered in the course of the MUCHFUSS (Massive Unseen Companions to Hot Faint Underluminous Stars from SDSS\footnote{Sloan Digital Sky Survey}) project, which orginally  aims at finding sdBs with compact companions like supermassive white dwarfs ($M>1.0\,M_{\rm \odot}$), neutron stars or black holes \cite{geier10c,geier10d}. The companion of SDSS\,J082053.53+000843.4 is the first unambiguously detected brown dwarf orbiting an sdB star. In the case of SDSS\,J162256.66+473051.1, which has the shortest orbital period of all known HW\,Vir systems ($\simeq0.075\,{\rm d}$), the companion may be a brown dwarf as well (see Schaffenroth et al. these proceedings). However, a detailed analysis of this system is necessary to prove this. 

Another promising way of finding HW\,Vir systems is provided by ongoing planetary transit surveys both from the ground and from space. Such binaries have been discovered with light curves from the NSVS \citep{wils07,for10}, the ASAS and the SWASP surveys (Schaffenroth et al. these proceedings). Most recently, \citet{oestensen10} reported the discovery of an HW\,Vir type star by the Kepler mission. 

In an ongoing project we are analysing light curves obtained by the CoRoT satellite \citep{rouan98,baglin06}. We used the AAOmega instrument to obtain medium resolution spectra of $\approx19\,000$ bright stars in the CoRoT fields. Our sample of 180 hot stars contains main sequence stars of A and B type, BHB stars as well as 25 candidate sdB stars. Five of the sdB candidates show eclipses. Follow-up time resolved spectroscopy will be obtained in order to derive orbital solutions and to constrain the atmospheric parameters of the stars better.


\section{Summary and outlook}

We have briefly reviewed the observational evidence for close substellar companions to hot subdwarf stars. Such objects can be easily detected by medium or high resolution spectroscopy with an RV accuracy at the ${\rm km\,s^{-1}}$-level. Eclipsing system of HW\,Vir type stick out because of their characteristic light curves. The best evidence that such objects are present and able to help forming sdBs is provided by the eclipsing system SDSS\,J082053.53+000843.4. Furthermore, several candidate systems have been discovered. The question whether substellar companions play a role in late stellar evolution in general and influence the formation process of hot subdwarf stars can be answered with yes. Future work is needed to explore the extent of this influence. Ongoing projects are RV surveys at high and medium resolution as well as light curve analyses with data coming from ground-based observatories and from space.


\begin{theacknowledgments}

A.T., S.G. and H.H. are supported by the Deutsche Forschungsgemeinschaft (DFG) through grants HE1356/45-1, HE1356/49-1, and HE1356/44-1, respectively. R.\O. acknowledges funding from the European Research Council under the European Community's Seventh Framework Programme (FP7/2007--2013)/ERC grant agreement N$^{\underline{\mathrm o}}$\,227224 ({\sc prosperity}), as well as from the Research Council of K.U.Leuven grant agreement GOA/2008/04. 

\end{theacknowledgments}



\bibliographystyle{aipprocl} 



\end{document}